\documentclass{aa}  
\usepackage{graphicx}
\usepackage{txfonts}
\usepackage[colorlinks=true,linkcolor=blue,urlcolor=blue,citecolor=blue,anchorcolor=blue]{hyperref}
\begin{document}


\title{Correlation between lithium abundances and ages of \\ solar twin stars
\thanks{Based on data products from observations made with ESO Telescopes at
the La Silla Paranal Observatory,
(observing programs 072.C-0488, 183.C-0972, 188.C-0265, and 0.88.C-0323).}}


   \author{Mar\'ilia Carlos
          \inst{1},
          Poul E. Nissen
          \inst{2}
          \and
          Jorge Mel\'endez\inst{1}
          }

   \institute{Universidade de São Paulo, IAG, Departamento de Astronomia, Rua do Matão 1226, Cidade Universitária,
05508-900 São Paulo, SP, Brazil\\
              \email{marilia.carlos@usp.br}
         \and
             Aarhus University, Stellar Astrophysics Centre, Department of Physics and Astronomy, Ny Munkegade 120, DK–8000 Aarhus C, Denmark\\
                         }

   \date{Received September 30, 2015; accepted December 18, 2015}

 
  \abstract
   {}
   {We want to determine the lithium abundances of solar twin stars as a function of
stellar age to provide constraints for stellar evolutions models and to investigate  whether there is
a connection between low Li abundance and  the occurrence of planets.
   }
   { For a sample of 21 solar twins observed with the HARPS spectrograph
at high spectral resolution (R $\simeq$ 115.000) and very high signal-to-noise ratio
(600 $\leq$ S/N $\leq$ 2400), precise lithium abundances were obtained by spectral synthesis of the
\ion{Li}{i} 6707.8 $\AA$ line and compared to stellar ages, masses, and metallicities determined
from a spectroscopic analysis of the same set of HARPS spectra.
  
  }
   { We show that for the large majority  of the solar twins there is a strong correlation between
lithium abundance and stellar age. As the age increases from 1 to 9\,Gyr, the Li abundance
decreases by a factor of $\sim$\,50.
The relation agrees fairly well with predictions from non-standard stellar evolution models of Li destruction at
the  bottom of the upper convection zone. Two stars deviate from the relation by having Li
abundances enhanced by a factor of $\sim$\,10, which may  be due to planet engulfment.
On the other hand, we find no indication of a link between planet hosting stars and
enhanced lithium depletion.
  }
   {}


   \maketitle

\section{Introduction}

Lithium is destroyed in the inner layers of  stars via proton capture ($^{7}Li(p, \alpha)\alpha$) 
at temperatures near to 2.5x10$^6$ K.  Because Li burning happens when the element
is transported to the innermost and hotter regions through convective motions in the star, 
Li abundance studies offer an excellent opportunity to understand the extent of the mixing 
processes within and below the stellar convective zone and, therefore, are important
 for constraining transport mechanisms in stars.

Since  the amount of lithium burning depends on the convective zone thickness, which depends on 
mass and metallicity, we can improve our understanding of the structure and evolution of a star 
by analyzing the lithium abundance dependence of these variables. It is important to note that 
the lithium content could also have other dependencies such as planets around a star; 
for instance, planetary formation could change the initial angular momentum of the star, 
which, according to \citet{takeda/10} and \citet{gonzalez/10},  increases  the lithium burning. 
Another alternative to an anomalous Li abundance is  planet engulfment,  which may 
lead to an increase in the Li abundance of a star \citep{montalban/02, sandquist/02}.

Although several factors  control the level of lithium burning, there is a debate regarding the 
lithium depletion in solar twin stars. There are two main interpretations. One is that the degree of
lithium depletion  primarily depends on stellar age, 
in the sense that older stars have lower Li abundances
than younger ones  as has been shown by \citet{baumann/10}, \citet{monroe/13}, \citet{melendeza/14},
and \citet{tucci/15}. This is supported by several models such as \citet{andrassy/15}, 
\citet{charbonnel/talon/05}, \citet{denissenkov/10}, \citet{donascimento/09}, and
\citet{xiong/deng/09}. The second interpretation is that enhanced Li depletion is
due to the presence of planets \citep{israelian/09, sousa/10, mena/14, figueira/14, gonzalez/15}
 and that the degree of lithium depletion does not depend on age for stars older
than $\sim 2$\,Gyr \citep{sousa/10, mena/14}.

In this paper we revisit the discussion  on lithium abundance depletion in 
solar twin stars and its correlations with stellar age, metallicity, mass, and planet occurrence
based on new, very precise Li abundances derived from HARPS spectra of
21 solar twins with ages between 0.7 and 8.8 Gyr. 

\section{Spectra and stellar parameters}

The solar twin stars included in this paper have been applied by \cite{nissen/15}
to study trends of abundance ratios as a function of stellar age and elemental condensation
temperature. The stars were selected from \cite{sousa/08} to have effective
temperatures ($T_{\rm eff}$),
surface gravities (log\,$g$), and metallicities ([Fe/H]) close to the solar values and
to have HARPS spectra \citep{mayor/03} with a signal-to-noise ratio $S/N \ge 600$
after combination of spectra available in the  ESO Science Archive.
Furthermore, a solar flux HARPS spectrum observed in reflected sunlight from the
asteroid Vesta is available. This spectrum has $S/N \sim 1200$  and was used in
a differential analysis of the stars relative to the Sun.
 
Details about how the HARPS spectra were normalized and used to determine stellar parameters
and abundances by a model atmosphere analysis of equivalent widths
are given in \cite{nissen/15}. Here we just mention that $T_{\rm eff}$ and log\,$g$
were determined by requesting that the iron abundances derived have no systematic
dependence on excitation and ionization potential of the lines applied.
The estimated errors of the parameters are $\sigma (T_{\rm eff}) = \pm 6$\,K and
$\sigma ({\rm log} g) = \pm 0.012$\,dex, and abundances were determined with a
typical precision of $\pm 0.01$\,dex. Such extremely high precision is obtainable
when high-resolution spectra of solar twin stars with $S/N > 500$ are analyzed
strictly differentially (line-by-line) relative to a spectrum of sunlight reflected
from an asteroid observed with the same spectrograph as the stars \citep{bedell/14}.
 
As seen from Table ~\ref{table:1}, the solar twin stars range from
5690 to 5870\,K in $T_{\rm eff}$, 4.25 to 4.50\,dex in log\,$g$,
and $-0.11$ to +0.11\,dex in [Fe/H]. Hence,
they lie in a region of the $T_{\rm eff}$ - log\,$g$ diagram,
where isochrones are not well separated \citep[see Fig. 5 in][]{nissen/15}.
Still, the small errors of $T_{\rm eff}$ and log\,$g$ make it possible to
derive precise ages by comparing with the Yonsei-Yale set of isochrones \citep{yi/01,
kim/02}. These ages are given in Table ~\ref{table:1} together with stellar masses 
estimated from the evolutionary tracks of \citet{yi/03}. A precision of
$\pm 0.01 M_{\odot}$ is estimated for the masses.
 
As seen from Table ~\ref{table:1}, the errors of the ages range from 0.4 to 0.8\,Gyr.
A comparison with ages derived by \cite{ramirez/14} from independent high $S/N$ spectra
of 14 of the stars shows good agreement within these
estimated errors \citep[see Fig. 6 in][]{nissen/15}. Furthermore, the abundance ratio [Y/Mg]
shows a  tight correlation with stellar age in a way that can be explained from the
nucleosynthesis of these elements, i.e., delayed production of yttrium in low-mass AGB stars
and prompt formation of magnesium in high mass, core collapse supernovae
\citep{nissen/15}. This correlation supports that the estimated age
errors are realistic  in a relative sense; the absolute ages for the youngest
and the oldest stars could be more uncertain.

\section{Analysis}

\begin{figure}
   \centering
   \includegraphics[width=\hsize]{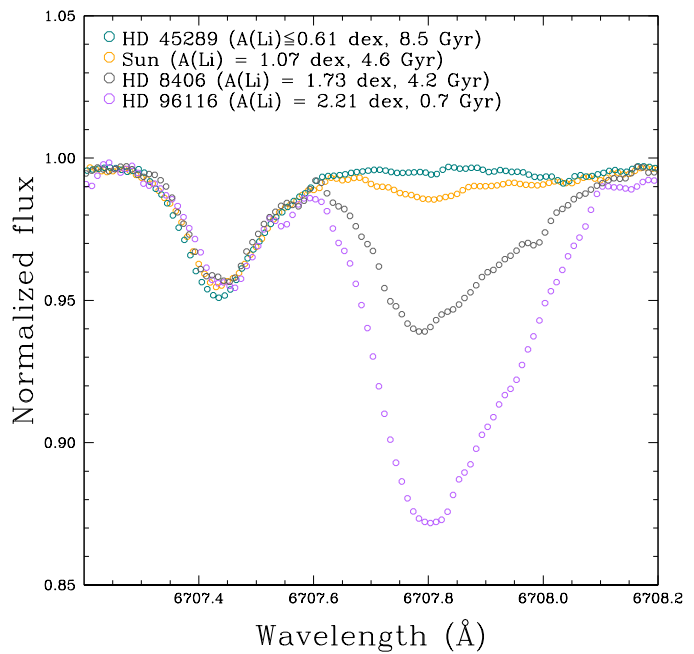}
      \caption{Observed spectra for four stars with different ages. Notice the weakening of the Li feature for increasing ages. 
              }
         \label{all}
   \end{figure}

An inspection of the observed spectra (Fig. \ref{all}) shows that there is a clear weakening 
of the Li feature with increasing ages.

In order to obtain the lithium abundances, first we had to estimate the 
macroturbulence and rotational velocity broadening. The macroturbulence velocity was
determined by the equation $v_{\rm macro,\star}=v_{\rm macro,\odot}+(T_{\rm eff}-5777)/486$\,km\,s$^{-1}$  as given 
in \citet{tucci/15}.  From our analysis of the HARPS solar spectrum we estimated $v_{\rm macro,\odot} = 3.2$ km\,s$^{-1}$, adopting $v\sin i_{\odot} = 1.9 $ km\,s$^{-1}$  . The projected rotational velocity was calculated 
by analyzing the line profiles of the Fe I 6027.050 $\AA$, 6093.644 $\AA$, 6151.618 $\AA$, 
6165.360 $\AA$, 6705.102 $\AA$, and Ni I 6767.772 $\AA$ lines. Notice that the instrumental broadening was also included.

\begin{figure}
   \centering
   \includegraphics[width=\hsize]{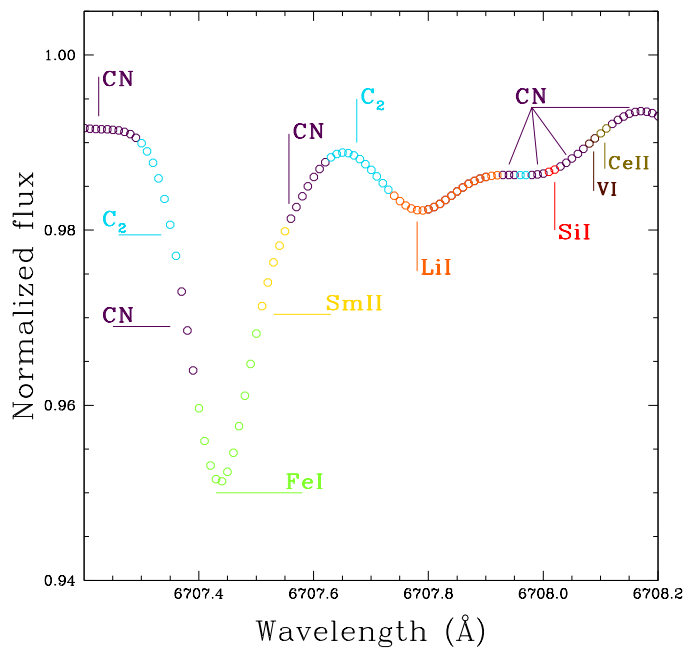}
      \caption{Synthetic solar spectrum considering different species in the region of the  6707.75 $\AA$ \ion{Li}{i} line. 
              }
         \label{sun}
   \end{figure}

\begin{figure}
   \centering
   \includegraphics[width=\hsize]{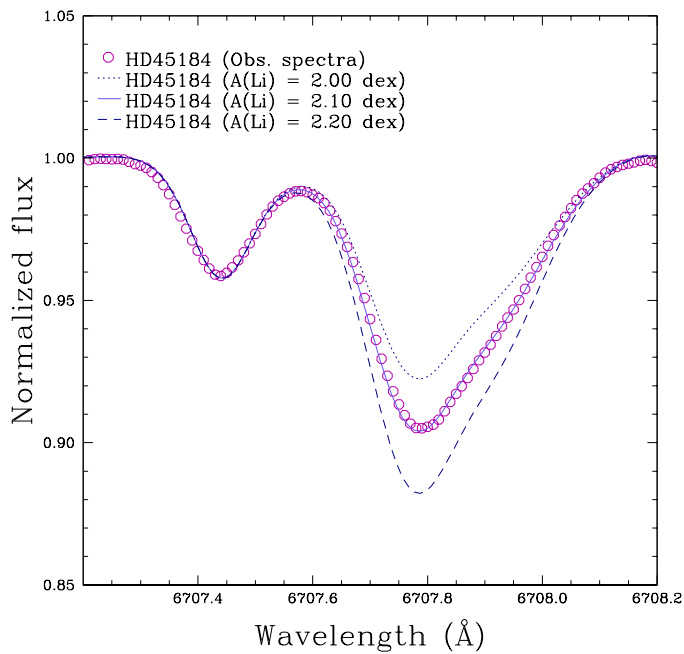}
      \caption{Open circles represent the observed spectra for HD 45184 in comparison with 
       three different synthetic spectra. 
              }
         \label{hd45184}
   \end{figure}

The lithium abundances were estimated by spectral synthesis in the region of the  6707.75 $\AA$ 
\ion{Li}{i} line using the July 2014 version of the 1D LTE code MOOG \citep{sneden/73}
and the Kurucz new grid of ATLAS9 model atmospheres \citep{castelli/04}. 
The line list adopted for the Li region is from \cite{melendez/12}, and includes blends 
from atomic and molecular (CN and C$_{2}$) lines; the dependence of the different species is shown in Fig. \ref{sun}. An example of spectral synthesis for
HD 45184 is shown in Fig. \ref{hd45184}, where it can be seen that the Li line profile 
changes very significantly for an abundance difference of 0.1 dex.

The abundances were derived assuming LTE and then NLTE results were obtained with the aid of 
the INSPECT database\footnote{www.inspect-stars.com (version 1.0).}, based on NLTE 
calculations by \citet{lind/09}. The errors were obtained taking into account uncertainties 
in the continuum setting,  the rms deviation of the observed line profile 
relative to the synthetic profile, and the stellar parameters. 
Both LTE and NLTE values of $A$(Li) = log($N_{\rm Li} / N_{\rm H})$ + 12 with their 
respective errors are shown in Table ~\ref{table:1}.

\begin{table*}
\caption{Li abundances, ages, masses and stellar parameters.}             
\label{table:1}      
\centering          
\begin{tabular}{l c c c c c c c r}     
\hline\hline       
\noalign{\smallskip}
Star &\multicolumn{1}{c}{\begin{tabular}[c]{@{}c@{}}$A$(Li) LTE\\(dex)\end{tabular}}&\multicolumn{1}{c}{\begin{tabular}[c]{@{}c@{}}$A$(Li) NLTE\\ (dex)\end{tabular}}&\multicolumn{1}{c}{\begin{tabular}[c]{@{}c@{}}Age\\ (Gyr)\end{tabular}}&\multicolumn{1}{c}{\begin{tabular}[c]{@{}c@{}}Error\\ (Gyr)\end{tabular}}&\multicolumn{1}{c}{\begin{tabular}[c]{@{}c@{}}Mass \\ (M$_{\odot}$) \end{tabular}} & \multicolumn{1}{c}{\begin{tabular}[c]{@{}c@{}} T$_{\rm eff}$ \\ (K) \end{tabular}}& \multicolumn{1}{c}{\begin{tabular}[c]{@{}c@{}} log \textit{g} \\ (dex) \end{tabular}}& \multicolumn{1}{c}{\begin{tabular}[c]{@{}c@{}}[Fe/H] \\ (dex)\end{tabular}}\\

\hline 
\noalign{\smallskip}
HD 2071&1.39$^{+0.02}_{-0.04}$&1.43$^{+0.02}_{-0.04}$&3.5&0.8&0.97&5724&4.490&$-$0.084\\
\noalign{\smallskip}
HD 8406&1.69$^{+0.01}_{-0.01}$&1.73$^{+0.01}_{-0.01}$&4.2&0.8&0.96&5730&4.479&$-$0.105\\
\noalign{\smallskip}
HD 20782\tablefootmark{a}&0.67$^{+0.08}_{-0.12}$&0.71$^{+0.08}_{-0.12}$&7.5&0.4&0.97&5776&4.345&$-$0.058\\
\noalign{\smallskip}
HD 27063&1.67$^{+0.02}_{-0.01}$&1.71$^{+0.02}_{-0.01}$&2.6&0.6&1.04&5779&4.469&0.064\\
\noalign{\smallskip}
HD 28471&0.86$^{+0.10}_{-0.09}$&0.90$^{+0.10}_{-0.09}$&7.0&0.4&0.97&5754&4.380&$-$0.054\\
\noalign{\smallskip}
HD 38277&1.56$^{+0.03}_{-0.03}$&1.58$^{+0.03}_{-0.03}$&7.3&0.4&1.01&5860&4.270&$-$0.070\\
\noalign{\smallskip}
HD 45184\tablefootmark{b}&2.08$^{+0.01}_{-0.02}$&2.10$^{+0.01}_{-0.02}$&2.7&0.5&1.06&5871&4.445&0.047\\
\noalign{\smallskip}
HD 45289\tablefootmark{$\ast$}&$\leq$0.57&$\leq$0.61&8.5&0.4&1.00&5718&4.284&$-$0.020\\
\noalign{\smallskip}
HD 71334&0.58$^{+0.14}_{-0.11}$&0.62$^{+0.14}_{-0.11}$&8.1&0.4&0.94&5701&4.374&$-$0.075\\
\noalign{\smallskip}
HD 78429&0.59$^{+0.11}_{-0.19}$&0.63$^{+0.11}_{-0.19}$&7.5&0.4&1.04&5756&4.272&0.078\\
\noalign{\smallskip}
HD 88084&0.96$^{+0.12}_{-0.06}$&1.00$^{+0.12}_{-0.06}$&6.0&0.6&0.96&5768&4.424&$-$0.091\\
\noalign{\smallskip}
HD 92719&1.88$^{+0.01}_{-0.01}$&1.90$^{+0.01}_{-0.01}$&2.7&0.6&0.99&5813&4.488&$-$0.112\\
\noalign{\smallskip}
HD 96116&2.19$^{+0.01}_{-0.01}$&2.21$^{+0.01}_{-0.01}$&0.7&0.7&1.05&5846&4.503&0.006\\
\noalign{\smallskip}
HD 96423&1.88$^{+0.01}_{-0.01}$&1.93$^{+0.01}_{-0.01}$&6.0&0.4&1.03&5714&4.359&0.113\\
\noalign{\smallskip}
HD 134664&2.08$^{+0.02}_{-0.01}$&2.11$^{+0.02}_{-0.01}$&2.3&0.5&1.07&5853&4.452&0.093\\
\noalign{\smallskip}
HD 146233&1.58$^{+0.02}_{-0.03}$&1.62$^{+0.02}_{-0.03}$&3.8&0.5&1.04&5809&4.434&0.046\\
\noalign{\smallskip}
HD 183658&1.24$^{+0.04}_{-0.07}$&1.28$^{+0.04}_{-0.07}$&5.0&0.5&1.03&5809&4.402&0.035\\
\noalign{\smallskip}
HD 208704&1.05$^{+0.05}_{-0.03}$&1.08$^{+0.05}_{-0.03}$&6.9&0.4&0.98&5828&4.346&$-$0.091\\
\noalign{\smallskip}
HD 210918\tablefootmark{$\ast$}&0.67$^{+0.06}_{-0.17}$&0.71$^{+0.06}_{-0.17}$&8.5&0.4&0.96&5748&4.319&$-$0.095\\
\noalign{\smallskip}
HD 220507\tablefootmark{$\ast$}&<0.50&<0.55&8.8&0.4&1.01&5690&4.247&0.013\\
\noalign{\smallskip}
HD 222582\tablefootmark{c}&0.89$^{+0.08}_{-0.05}$&0.93$^{+0.08}_{-0.05}$&6.5&0.4&1.01&5784&4.361&$-$0.004\\
\noalign{\smallskip}
Sun&1.03$^{+0.03}_{-0.02}$&1.07$^{+0.03}_{-0.02}$&4.6&--&1.00&5777&4.438&0.000\\
\noalign{\smallskip}
\hline                  
\end{tabular}
\tablefoot{
\tablefoottext{$\ast$}{$\alpha$-enhanced star.}
\tablefoottext{a}{Detected planet with 1.8 $M_{\rm Jup}$ \citep{jones/06}.}
\tablefoottext{b}{Detected planet with 0.04 $M_{\rm Jup}$ \citep{mayor/11}.}
\tablefoottext{c}{Detected planet with 7.8 $M_{\rm Jup}$ \citep{butler/06}.}
}
\end{table*}

\begin{figure}
   \centering
   \includegraphics[width=\hsize]{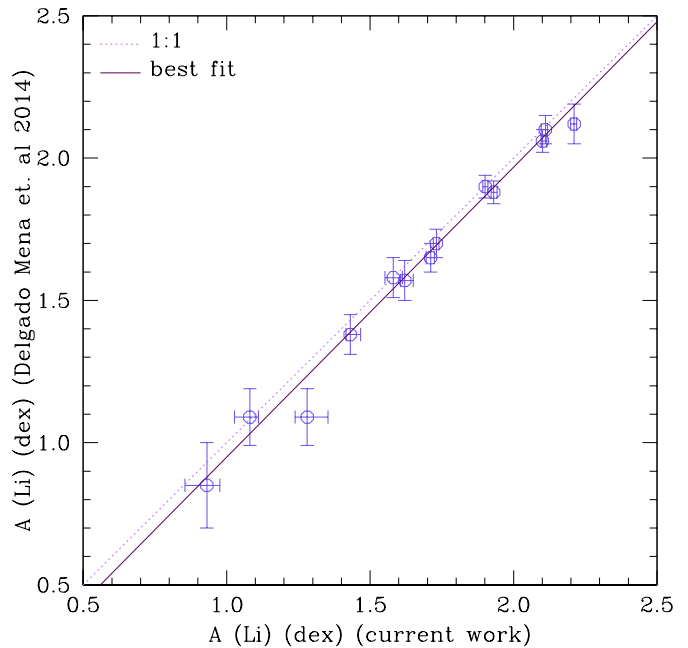}
      \caption{LTE Li abundances for our sample compared with LTE data from the work of \citet{mena/14}.
              }
         \label{compare}
   \end{figure}

It is important to note that our results in LTE compare well with the large study by
\citet{mena/14}, as shown in  Fig. ~\ref{compare}. Compared to previous works 
on solar twins
(\citet{monroe/13} for 18 Sco and HIP 102152, \citet{melendez/12} for HIP 56948, 
\citet{melendeza/14} for HIP 114328\footnote{A new better spectrum of HIP 114328 has been recently acquired by one of the authors (J.M.). The stellar parameters are T$_{\rm eff}=5763\pm12$ K, log $g$ = 4.33 $\pm$ 0.04 dex and [Fe/H]=-0.050 $\pm$ 0.014 dex, resulting in an age of $7.6^{+0.6}_{-0.9}$ Gyr and mass equal to 1.00 $\pm$ 0.01 M$_{\odot}$. Considering these new results the star fits better now the Li-age trend.} and \citet{ramirez/11} for 16 Cyg A and 16 Cyg B), 
our results, now in NLTE, follow the same lithium age trend, 
as illustrated in Fig. ~\ref{liage1}. The effective temperatures and
surface gravities of the stars in these papers were determined in the same way
as in \citet{nissen/15}, i.e., by a differential analysis of Fe lines relative to the Sun,
and ages were also obtained by interpolating between Yonsei-Yale isochrones. Hence, we
expect that the Li abundances and ages are on the same system as in the present paper.

\section{Discussion}

\begin{figure}
   \centering
   \includegraphics[width=\hsize]{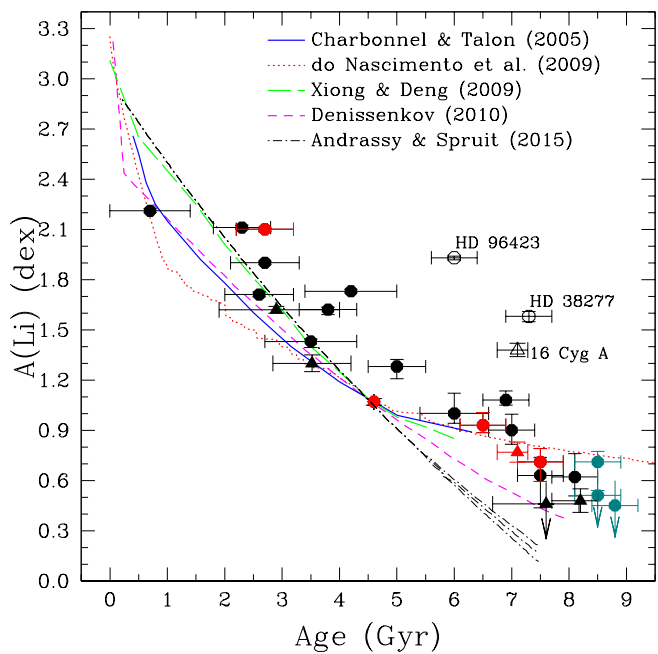}
      \caption{Connection between stellar ages and NLTE lithium abundances for our current sample
(circles) and some previous results (triangles) referenced in the text. 
Teal blue circles indicate alpha-enhanced stars and red symbols refer to stars hosting planets. 
The models of Li depletion were normalized to the solar Li abundance. 
In some cases, the lithium abundance errors are smaller than the points.
              }
         \label{liage1}
   \end{figure}

The results for the whole sample can be seen in Fig. ~\ref{liage1}, where we notice a 
strong connection between lithium abundance and stellar age. 16 Cyg A 
and two stars in our sample, namely HD 96423 and HD 38277, have higher  abundances than expected
in comparison to both models and the other stars. A possible explanation for this Li enhancement 
could be planet engulfment as described in \citet{montalban/02} and \citet{sandquist/02}. 

Another interesting result is that the 
$\alpha$-enhanced stars (represented by teal blue filled circles in Fig. ~\ref{liage1}) 
fall on the same Li-age sequence as the thin disk solar twins. Two of these 
$\alpha$-enhanced stars (HD\,45289 and HD\,210918) have thick-disk kinematics,
while the third one (HD\,220507) has thin-disk kinematics like the rest of the stars
\citep{nissen/15}, but this latter star has a chemical abundance pattern consistent with the other two thick disk stars.

Considering all stars in our sample for which a Li abundance has been derived
(i.e., excluding stars with only an upper limit), 
the Li-age correlation has about a 10$\sigma$ significance; 
we performed a Spearman test and found a Spearman rank coefficient $r_{s}=-0.89$ and  
a probability of  $10^{-8}$ of our results arising by pure chance. Therefore, even when 
the two deviant stars are considered the correlation between Li and age is very strong. 

Excluding the anomalous stars HD 96423 and HD 38277, the Li-age correlation
 for 17 stars in our sample that have firm detections of the Li line is at the
23$\sigma$ significance level and has a very small probability of not being real ($10^{-11}$) 
with a Spearman rank coefficient $r_{s}=-0.96$, meaning an almost perfect Spearman correlation. 
This is in contrast to the poor correlation found by \cite{mena/14} (their Fig. 8). Notice, 
however, that they include stars with masses significantly lower and higher than the Sun's 
and  they also include stars in a broader metallicity range. The wider interval in both 
metallicity and mass apparently blurs the strong correlation that we find.

In Fig. \ref{liage1} we compare our Li-age results with Li depletion models for 
a one solar mass star with solar metallicity
\citep{xiong/deng/09, donascimento/09, charbonnel/talon/05, denissenkov/10, andrassy/15} 
that take into account physics not included in the standard solar model. Different transport 
mechanisms are explored in these models, including meridional circulation, diffusion 
(gravitational settling and radiative acceleration), turbulence, gravity waves, convective 
overshooting, and convective settling. Overall, all models have a reasonable qualitative 
agreement with the data, as expected because they were calibrated using the Sun or earlier 
observations of solar twins.  With the data presented here, the models above
could be improved. However, such refinements should take into
account that the initial Li abundance of stars probably have increased in time
due to Galactic evolution, i.e., Li production in AGB stars and novae \citep{chen/01, 
romano/01, lambert/04, ramirez/12}.  The models assume an initial
Li abundance of $A$(Li) = 3.3 as in meteorites, but according to the chemical
evolution model of \citet{izzo/15}, the initial Li abundance has increased from about $A$(Li) = 2.8
for the oldest disk stars to about $A$(Li) = 3.6 for the youngest. Taking this into
account would make the $A$(Li) - age relation predicted by the models steeper than
shown in Fig.~\ref{liage1}.

 For models including rotational induced
mixing, variations of the initial rotational velocity cause a considerable scatter
in $A$(Li) at a given age; according to \citet{charbonnel/talon/05} a change in 
$V_{\rm rot, init.}$ from 15\,km\,s$^{-1}$  to 110\,km\,s$^{-1}$ decreases $A$(Li) by about 1\,dex
for the oldest stars. Differences in the initial rotational velocity
may, therefore, contribute to the scatter seen in Fig.~\ref{liage1}, although
very large differences in $V_{\rm rot, init.}$ are needed to explain the outliers.

An alternative explanation for enhanced lithium depletion is the presence of planets.
\citet{israelian/09} suggest that planet hosting stars have less lithium than stars without planets, but 
\citet{baumann/10} contest this result claiming that the sample of the former work
is biased in metallicity and includes stars of different evolutionary stages. 
In agreement with \cite{baumann/10}, Fig. \ref{liage1} shows a lack of connection between 
lithium depletion and presence of planets. There is no significant segregation between stars 
without planets (black filled symbols) and  planet hosting stars (red filled symbols) in this figure.

\begin{figure}
   \centering
   \includegraphics[width=\hsize]{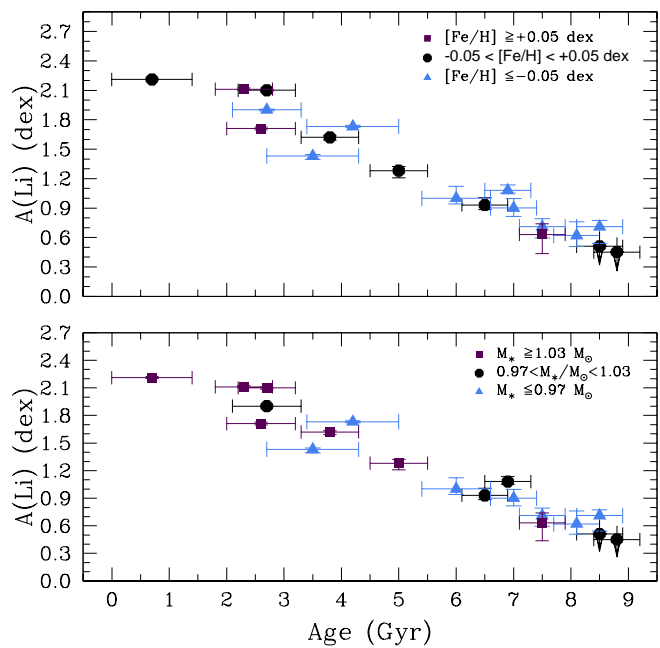}
      \caption{Connection between stellar ages and lithium abundances with different symbols
        for three metallicity groups (top) and three mass groups (bottom). 
              }
         \label{liage3}
   \end{figure}

As shown in Fig. \ref{liage3}, where the stars in our sample have been divided into three
metallicity groups (upper panel) and three mass groups (lower panel), there is no
obvious indication of a dependence of lithium depletion on metallicity and mass
at a given age. An unweighted linear least squares fit to $A$(Li) as a function
of age (in Gyr) for stars with a detected Li line (including the Sun) leads to 
\begin{eqnarray}
A({\rm Li}) = 2.437 \, (\pm 0.098) - 0.224 \, (\pm 0.018) \,\, Age
\end{eqnarray}
Fitting the residuals of $A$(Li) relative to this equation as a function of [Fe/H] and mass
results in the relation
\begin{eqnarray}
\Delta\,A({\rm Li}) = -0.05 \, -2.17 \, (\pm 1.30) \,\, {\rm [Fe/H]} \nonumber \\
      + 3.55 \, (\pm 2.25) \,\, (M / M_{\odot} - 1.0)
\end{eqnarray}
Although, the coefficients have the expected sign, i.e., increasing Li depletion 
with increasing metallicity and decreasing mass \citep[e.g.,][]{castro/09}, 
the dependence is hardly significant, i.e., a 1.7$\sigma$ correlation with
metallicity and a 1.6$\sigma$ correlation with mass. Furthermore, the mass dependence
is smaller than predicted from models \citep{xiong/deng/09, castro/09}, and also 
smaller than inferred from Li abundance measurements for stars in the open cluster M67,
which suggests a mass coefficient on the order of 10 for $\sim\!  1 M_{\odot}$ stars
\citep{pace/12}. Apparently, the short mass range for our sample
(from  0.94\,$M_{\odot}$ to 1.07\,$M_{\odot}$) in combination with 
errors in age, mass, and Li abundance make it difficult to estimate
the dependence of Li abundance on mass based on our sample. 
We note, however, that according to our results, it is unlikely that the large difference
in $A$(Li) between 16 Cyg A and 16 Cyg B is due to the difference in mass (0.05 $M_{\odot}$)
as suggested by \citet{ramirez/11}. Instead, the high Li abundance in 16 Cyg A 
may be due to a recent accretion of a Jupiter  sized planet, which is consistent with the
fact that no such planet has been detected for 16 Cyg A, whereas 16 Cyg B  has
a  giant planet with a minimum mass of 1.5\,$M_{\rm Jup}$ \citep{cochran/97}. A related evidence supporting the planet engulfment scenario could be the enhanced abundances for all elements in 16 Cyg A relative to 16 Cyg B \citep{ramirez/11,tucci/14}, and in particular for the refratory elements \citep{tucci/14}.

\section{Conclusions}

Based on HARPS spectra with very high signal-to-noise ($S/N \ge 600$) we  were able 
to determine high-precision lithium abundances for 21 solar twin stars for which isochrone 
ages have been estimated from the $T_{\rm eff}$ - log\,$g$ diagram. Excluding two outliers,  
we see a strong correlation between stellar ages and lithium abundances, 
with a 23$\sigma$ significance and a Spearman test coefficient of $r_{s}=-0.96$.  In addition, five solar twins for which high-precision lithium abundances and ages  were previously derived \citep{monroe/13, melendez/12, melendeza/14, ramirez/11} and the Sun itself fit the relation very well. 
Over an age range from 1 to 9\,Gyr, the Li abundance decreases by almost 2 dex, which
we interpret as being due to gradual destruction of lithium near the bottom of the outer
convection zone. Thus, the new data can be used to constrain non-standard models of Li depletion, 
and therefore to understand better the transport mechanisms inside stars.

Two stars deviate from the main trend between $A$(Li) and age by having Li abundances
enhanced by a factor of $\sim$\,10. A similar enhancement of Li is present for 16 Cyg A 
relative to 16 Cyg B   even though the two stars have the same age
and nearly the same mass and metallicity. We suggest that these enhancements of Li are connected
to  the accretion of Jupiter  sized planets.  

We find no evidence for a link between planet hosting stars and enhanced Li destruction
as suggested in some recent works. Three stars with planets of different masses in our sample,
16 Cyg B with a Jupiter  sized planet \citep{ramirez/11}, and the Sun show no systematic deviation
from the $A$(Li)-age relation for stars without planets.

\begin{acknowledgements}
The referee is thanked for important comments,  that helped to improve this paper.
M.C. would like to acknowledge support from CAPES. P.E.N. acknowledges support from 
the Stellar Astrophysics Centre funded by The
Danish National Research Foundation (Grant agreement no.: DNRF106) 
and from the ASTERISK project (ASTERoseismic Investigations with SONG and Kepler)
funded by the European Research Council (Grant agreement no.: 267864). 
J.M. would like to acknowledge support from FAPESP
(2012/24392-2) and CNPq (Bolsa de Produtividade).

\end{acknowledgements}

\nocite{*}

\bibliographystyle{aa}
\bibliography{./mybib}

\end{document}